%% file: main.tex
\documentclass[10pt,twocolumn,letterpaper]{article}
\pdfoutput=1

\usepackage{cvpr}              

\usepackage[accsupp]{axessibility}  

\usepackage{graphicx}
\usepackage{amsmath}
\usepackage{amssymb}
\usepackage{booktabs}
\usepackage{subcaption}
\usepackage{multirow}

\usepackage[pagebackref,breaklinks,colorlinks]{hyperref}

\usepackage[capitalize]{cleveref}
\crefname{section}{Sec.}{Secs.}
\Crefname{section}{Section}{Sections}
\Crefname{table}{Table}{Tables}
\crefname{table}{Tab.}{Tabs.}

\def\cvprPaperID{5998} 
\def\confName{CVPR}
\def\confYear{2022}

\begin{document}

\makeatletter
\newcommand{\printfnsymbol}[1]{%
  \textsuperscript{\@fnsymbol{#1}}%
}
\makeatother

\title{RIDDLE: Lidar Data Compression with Range Image Deep Delta Encoding}

\author{Xuanyu Zhou\thanks{equal contribution}~~~~~Charles R. Qi\printfnsymbol{1}~~~~~Yin Zhou~~~~~Dragomir Anguelov
\vspace{0.2cm}\\Waymo LLC}

\maketitle

\input{tex/abstract}

\section{Introduction}
\input{tex/intro}

\section{Related Work}
\input{tex/related_work}

\section{Problem Formulation}
\input{tex/problem}

\section{Range Image Deep Delta Encoding}
\input{tex/method}

\section{Experiments}
\input{tex/exp}

\section{Conclusion}
\input{tex/conclusion}

{\small
\bibliographystyle{ieee_fullname}
\bibliography{ref}
}

\newpage\phantom{blabla}

\newpage

\appendix
\section*{Supplementary}

\input{tex/appendix}

\newpage

\end{document}

%% file: tex/abstract.tex
\begin{abstract}

Lidars are depth measuring sensors widely used in autonomous driving and augmented reality. However, the large volume of data produced by lidars can lead to high costs in data storage and transmission.
While lidar data can be represented as two interchangeable representations: 3D point clouds and range images, most previous work focus on compressing the generic 3D point clouds. In this work, we show that directly compressing the range images can leverage the lidar scanning pattern, compared to compressing the unprojected point clouds. We propose a novel data-driven range image compression algorithm, named RIDDLE (Range Image Deep DeLta Encoding). At its core is a deep model that predicts the next pixel value in a raster scanning order, based on contextual laser shots from both the current and past scans (represented as a 4D point cloud of spherical coordinates and time). The deltas between predictions and original values can then be compressed by entropy encoding.
Evaluated on the Waymo Open Dataset and KITTI, our method demonstrates significant improvement in the compression rate (under the same distortion) compared to widely used point cloud and range image compression algorithms as well as recent deep methods.

\end{abstract}

%% file: tex/intro.tex

Lidar (or LiDAR, short for light detection and ranging) sensors are commonly used in applications that require 3D scene understanding such as autonomous driving and augmented reality.
However, with the growing resolution of lidars, storing and transmitting large volumes of sequential lidar data become a challenge. There is a strong need to develop effective algorithms for lidar data compression.

While the measurements of a lidar scan are often used as a 3D point cloud, the raw lidar data can be represented as a more structured format: a range image, where each pixel corresponds to a laser shot, each row represents shots from the same laser, each column represents shots at a specific azimuth rotation angle.
Given the lidar scanning mechanism (directions of the lasers) and sensor poses (6D poses in the global coordinate at the timestamp of every shot), a range image and its corresponding point cloud can be converted interchangeably and losslessly. By organizing the points in a range image, instead of storing the \emph{three}-dimensional coordinates of the points, we can just store \emph{one}-dimensional ranges (around 3x saving in storage). Given this observation, in contrast to previous works that focus on compressing 3D point clouds~\cite{huang2020octsqueeze, biswas2020muscle, que2021voxelcontext}, we propose to directly compress range images to leverage the lidar scanning patterns.

As range images are in the image format, naturally we can apply existing compression methods for optical images (RGB or grayscale); however, those methods have their limitations. For example, the PNG format is often used to compress depth images in indoor datasets~\cite{song2015sun, 2017arXiv170201105A, dai2017scannet}, where the depth value are normalized and quantized to 16-bit integers and compressed losslessly.
While PNG also applies to compress lidar range images, it is not data-driven and does not use temporal information. There are also attempts to use auto-encoder networks~\cite{tu2019point} to lossily compress range images by storing the bottleneck layer output. However, as range values often have a much wider distribution than RGB colors, it is challenging to learn an accurate reconstruction, especially at the object boundaries.


In this work, we propose \emph{RIDDLE} (Range Image Deep DeLta Encoding), a data-driven algorithm to compress range images with predictive neural networks (Fig.~\ref{fig:pipeline}).
Our method is inspired by the use of delta encoding in PNG image compression. However, instead of simply computing a difference between close-by pixels, we adopt a deep model to predict the pixel value from context pixels. 
The deep model takes a local patch of the decoded range image and predicts the attributes of the next pixel in a raster-scanning order (a similar process to the sequential image decoder PixelCNN~\cite{van2016pixel}).
We can then entropy encode the residuals between the predicted values and the original values to achieve \emph{lossless} compression under a chosen quantization rate. In this scheme, the more accurate the prediction is, the smaller the entropy of the residuals are -- improving the compression rate is equivalent to developing a more accurate predictive model.

What is unique in our model design is that we represent local image patches as point clouds in the \emph{spherical} coordinates (with azimuth, elevation and range values) to reflect the non-uniform ray angles of each shot (or pixel), which lifts the 2D pixels to 3D point clouds. By further lifting the 3D points to 4D with a timestamp channel, we can unify the way we represent context pixels/points from both the \emph{current} and \emph{history} scans.
Since our model directly takes in point clouds, neither interpolation (to the image grid) nor image cropping (projected points from history frames may span different image regions) is needed. On the other hand, as to the model output formulation, instead of directly regressing the pixel values (which is often multi-modal), we treat each pixel in the input patch as an anchor and predict a confidence score as well as a residual value per anchor.

 Evaluated on the large-scale Waymo Open Dataset (WOD)~\cite{sun2020scalability}, we show that our method reduces the bitrate by more than 65\% for the same distortion (measured using the point-to-point Chamfer distance) or reducing more than 85\% distortion for the same bitrate, compared to the MPEG standard compression method G-PCC~\cite{graziosi2020overview} while also significantly outperforming other baselines like Draco~\cite{Draco} and PNG. On the KITTI dataset~\cite{geiger2013vision}, we compare with prior art deep compression methods (using octrees) and show our method has a clear advantage over them, thanks to its use of the range image representation and the accurate prediction model. We also evaluate the impact of compression on downstream perception tasks such as 3D object detection and provide extensive ablation studies to validate our design choices.

%% file: tex/related_work.tex


\paragraph{Point cloud compression}
As 3D applications rise, recent years have seen an increasing number of algorithms proposed for point cloud compression. One family of the methods uses octrees to represent and compress quantized point clouds~\cite{botsch2002efficient, devillers2000geometric, schnabel2006octree}. The Motion Picture Experts Group (MPEG) has released a related point cloud compression (PCC) standard, called geometry-based PCC (G-PCC)~\cite{graziosi2020overview}, using the octree structure and various ways to predict the next-level content. More recently, Octsqueeze~\cite{huang2020octsqueeze} was proposed to use a neural network as a conditional entropy model to estimate the octree occupancy symbols, and MuSCLE~\cite{biswas2020muscle} extends it by including temporal prior from previous frames. VoxelContextNet~\cite{que2021voxelcontext} further leverages the voxel context for the octree structure prediction. These neural network-based methods consistently show improvements over G-PCC which uses hand-crafted entropy models. While the octree-based methods are flexible to model arbitrary point clouds (from either a lidar sensor or multi-view reconstruction), they do not make use of the point distribution patterns in lidar range images.

As a lidar point cloud can be represented as a range image, image-based compression methods can be adapted for its compression. For example, \cite{houshiar20153d, beek2019image, adaprad} applied traditional image compression methods such as JPEG, PNG and TIFF to compress the range images. A sequence of range images could be seen as a video, and video-based compression method like H.264 was applied to compress lidar sequences~\cite{6943095}. MPEG also proposed a PCC (V-PCC) standard that compresses dynamic point clouds via HEVC video codex~\cite{graziosi2020overview}. Our work extends them to leverage deep models and delta encoding to compress range images.

Auto-encoders have been used to achieve lossy compression of point clouds. \cite{yan2019deep, wiesmann2021deep} proposed to train an encoder-decoder point cloud reconstruction network and entropy encode the bottleneck layer as the compressed data. Similarly, \cite{tu2019point} trained an auto-encoder to reconstruct range images and compress the bottleneck vectors. While these methods may achieve high compression rates, the reconstructed point clouds could have strong artifacts, especially at the object boundaries resulting in unbounded errors in the lossy compression scheme.


\paragraph{Learned image and video compression}
Image and video compression are well-studied fields with many standards (for example: PNG, JPEG, TIFF for images, H.264 and HEVC for videos). Among them, PNG is highly related to our work as it uses lossless image compression using delta encoding.
With the popularity of deep convolutional neural networks for image understanding, deep model-based image and video compression have also been widely explored~\cite{balle2016end, toderici2017full, balle2018variational, townsend2019hilloc, mentzer2019practical, ma2019image}. Many of them leverage an encoder-decoder neural network (for example, a variational auto-encoder~\cite{balle2016end}) for the compressing (encoding the image to a latent vector) and decompressing (decode/generate the image from the vector).
For the decoding architectures, sequential models such as PixelCNN \cite{oord2016conditional} and PixelRNN \cite{van2016pixel} inspired our predictive model design.

%% file: tex/problem.tex
\begin{figure}
    \centering
    \includegraphics[width=\linewidth]{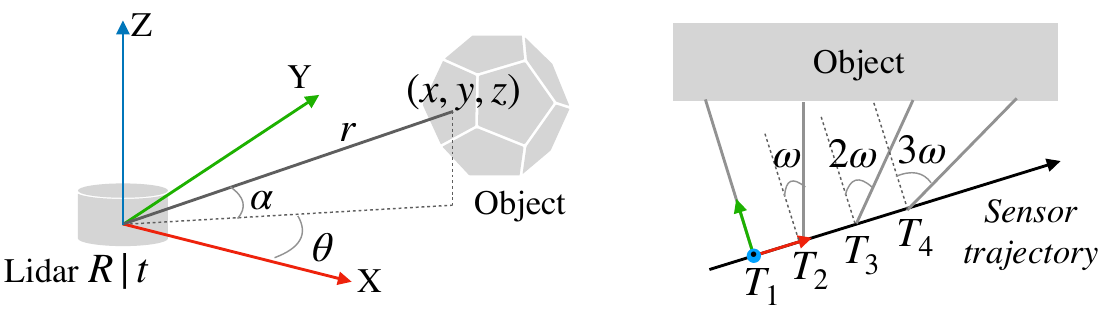}
    \caption{\textbf{Illustration of laser shots.} \emph{Left: } A single laser shot. \emph{Right: } Laser shots across time (in a bird's eye view). We show four consecutive laser shots (with delta azimuth angle $\omega$) that measure the ranges from the (moving) sensor to the object. To convert the range values to a point cloud, we need to know the ranges, the shot angles, as well as the sensor poses at each shot.}
    \label{fig:laser}
\end{figure}

\label{sec:problem}
For most lidar sensors, one scan can be interchangeably represented as either a point cloud $P \in \mathbb{R}^{N \times C}$ or a range image $I \in \mathbb{R}^{H \times W \times C}$, where N is the number of points, H and W are height and width of the range image (H is the number of laser beams in the lidar and W is the number of shots per laser per frame), C is the feature dimension for each point. Each valid pixel in the range image represents a laser shot corresponding to one point in the point cloud. The channels include the range value and other attributes such as reflection intensity. The conversion rule between a point cloud and a range image depends on the laser scanning mechanism (the laser shot azimuth and elevation angles) as well as the sensor poses (the 6D pose of the laser sensor at the time of each laser shot), as illustrated in Fig.~\ref{fig:laser}.

Specifically, in a range image $I$, given a pixel location $(i, j)$ (which maps to a specific laser shot angle) and its range value, we get a laser measurement $(r, \theta, \alpha)$ where $r$ is the range value, $\theta$ (azimuth or yaw) and $\alpha$ (elevation or pitch) are the shot angles relative to the lidar sensor coordinate. The measurement can be converted to a point $p$ in the sensor coordinate by:

\begin{equation}
\label{eq:unprojection}
   p = (x, y, z) = (r \cos \alpha \cos \theta, r \cos \alpha \sin \theta, r \sin \alpha)
\end{equation}

As at the time of each laser shot, the sensor pose $[R | t]$ (rotation and translation in the global coordinate) can be different (Fig.~\ref{fig:laser}). To aggregate the shots into a point cloud, we need to convert the points to a shared global coordinate system to get the point set $P = \{ R_i p_i^T + t_i\}, i = 1, ..., N$ where $i$ is the index of the laser shot in a scan/range image.

Reversely, given the point cloud $P$ of a scan (in the global coordinate), to convert it to the range image, we first need to transform \emph{each point} to the sensor coordinate corresponding to its time of the shot. Then, we can easily get the $(r, \theta, \alpha)$ by the reverse process of Eq.~\ref{eq:unprojection}, which then maps back to the row and column indices.

For our lidar range image compression, we first quantize the range image $I$ by rounding its pixel values to a predetermined quantization precision. Then our goal is to compress the quantized range image $I'$ to a bitstream $b \in [0, 1]^n$ (with an $n$ as small as possible), which can later be decompressed into the exact quantized range image $I'$. It is lossy with respective to the raw range image but \emph{lossless} regarding the quantized range image.

Note that for calibrated lidars such as the ones used in the Waymo Open Dataset~\cite{sun2020scalability}, each pixel in the range image corresponds to a fixed shot angle $(\theta, \alpha)$ for the same lidar, so the angles do not need to be stored for the compression~\footnote{For the main lidars used in WOD, pixel elevations are determined by the laser beam inclinations (64 numbers) and azimuths can be calculated based on uniform azimuth rotation. For other lidars such as Velodyne HDL-64, azimuth rotation angles are not uniform and need to be stored (one number for each column, costing only $\sim 0.1$Kb per frame)~\cite{tu2016compressing}.}. Besides, as sensor poses are often stored separately from range images and are shared with other modules (such as localization), we do not need to store sensor poses either. Only the range image needs to be compressed.

%% file: tex/method.tex


\begin{figure}[t!]
    \centering
    \includegraphics[width=\linewidth]{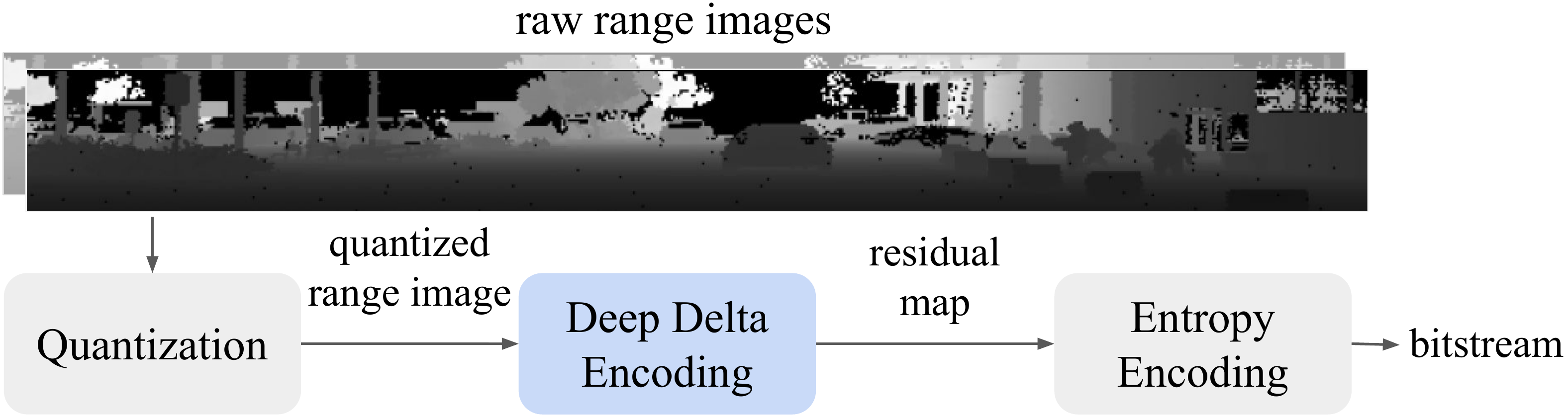}
    \caption{\textbf{The deep delta encoding pipeline for lidar range image compression.} Given a lidar range image, we first quantize the attribute values and then run inference of the predictive model on the quantized range image to derive residuals. Finally we use entropy encoders to compress the residuals to a bitstream. }
    \label{fig:pipeline}
\end{figure}

We first describe our overall compression pipeline in Sec.~\ref{sec:approach:pipeline}, then dive deep into the design of our prediction model in Sec.~\ref{sec:approach:network}, and finally describe how we entropy encode the residuals in Sec.~\ref{section:approach:entropy}.

\subsection{Pipeline Overview}
\label{sec:approach:pipeline}
As shown in Fig.~\ref{fig:pipeline}, the input to our compression pipeline is a raw range image. First, we quantize the range image with a certain quantization precision (this allows us to store the deltas as discrete symbols).
Next, the core part of the pipeline is the deep delta encoding. We train a deep model to predict the next pixel value in a raster scanning order. We then save the \emph{delta} between the prediction (quantized) and the original (quantized) pixel value instead of saving the original pixel value. As the deltas are smaller and more concentrated in distribution than the original pixel values, they can be compressed more effectively.
At the last step, the deltas (or the residual map) are entropy encoded to a compressed bitstream.

\subsection{Deep Delta Encoding}
\label{sec:approach:network}

Commonly used delta encoding adopts a linear prediction model to estimate the pixel values. In its simplest form, to predict a pixel $I_{i, j}$ at the $i$-th row and $j$-th column, its left pixel $I_{i, j-1}$ is used as the prediction. Other linear filters of left, up and nearby pixels can also be used. The \emph{delta} between the prediction and the original pixel value is stored to be compressed. In our work, we propose to train a deep neural network to predict the pixel values and show that it can achieve significant improvement in prediction accuracy and compression rate. Next, we first introduce our model in its intra-prediction format (only using the information from the current frame/scan for the prediction) and then describe how we extend it to take temporal input from history scans. Please see the supplementary for more details on the model architecture, the losses and the training process.

\begin{figure*}[t!]
    \centering
    \includegraphics[width=0.8\linewidth]{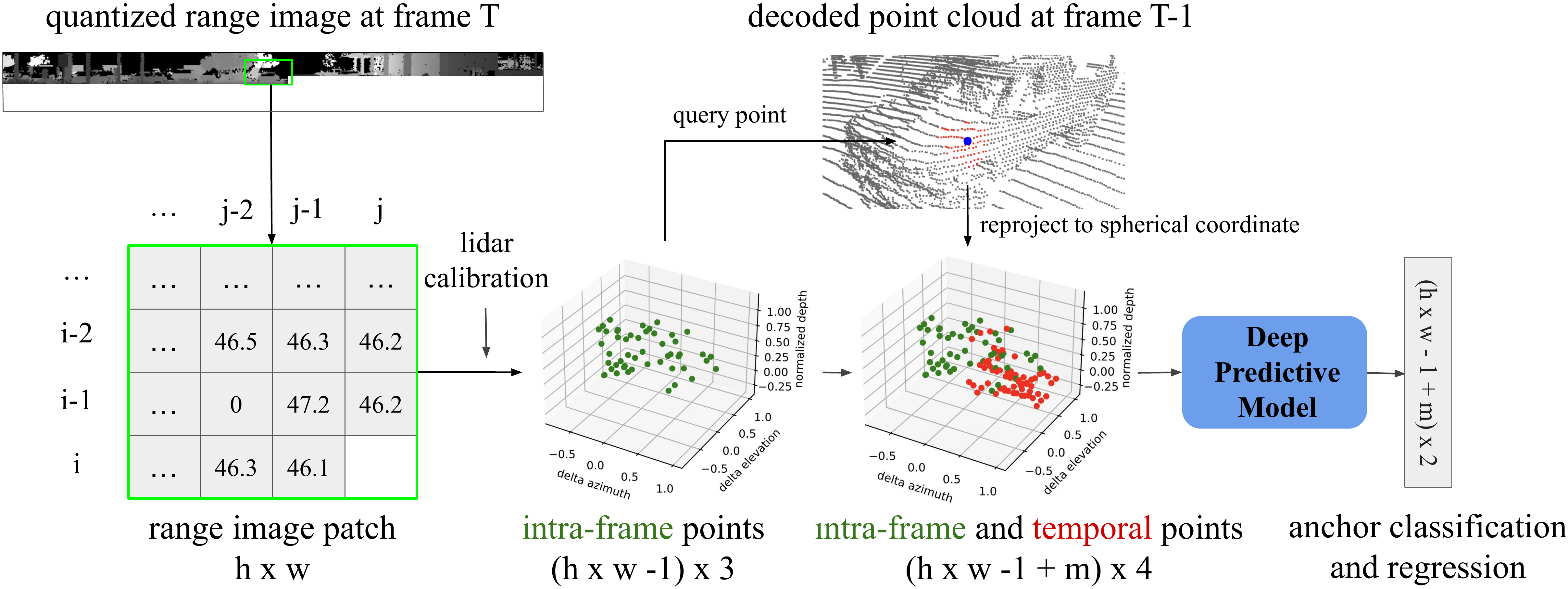}
    \caption{\textbf{The deep prediction model.} Given a range image patch from frame T with quantized attribute values (e.g. range), we lift pixels to the spherical coordinate with azimuth and elevation angles from lidar calibration. To leverage context points from the past frame T-1, a query point is generated to find neighbors among points at frame T-1. Those neighbor points are then projected to the spherical coordinate of the pixel to be predicted. Our predictor takes the union of the intra-frame and temporal context points and predicts the attribute of the pixel $(i, j)$ with anchor classification and regression (with each input point as an anchor).}
    \label{fig:network}
\end{figure*}

\paragraph{Intra-frame Prediction Model}
Formally, the network models the conditional probability of the $k$-th pixel value (in the raster scanning order) conditioned on the quantized pixel values before $k$: $p(I_{k};\Theta) = p(I_{k} | \{I'_{k-1}, ..., I'_{1}\}; \Theta)$, where $\Theta$ are the network weights, $I'$ is the quantized range image and $I$ is the unquantized raw range image. Empirically, as shown in Fig.~\ref{fig:network}, instead of using the entire past context (e.g. with a RNN model), we can use local image patch of shape $h \times w$ as the context to predict the bottom right pixel of the patch, similar to the idea of the sequential image decoder PixelCNN~\cite{oord2016conditional}.

Although the input to our network is an image patch, it is quite different from a typical RGB one. The relations of the range image pixels depend on the location of the patch and even the calibration of a specific lidar because the laser shot angles are often non-uniformly distributed. This is even more prominent in the inter-frame prediction when we re-project the points from history scans to the coordinate of the current shot. Therefore, we augment the range image with two extra channels: the delta azimuth and delta elevation angles relative to the angles of the to-be-predicted pixel, which lifts the 2D pixels to the 3D spherical coordinate. Furthermore, as range prediction is a geometry estimation problem, we found that empirically, using a 3D deep learning model such as PointNet~\cite{qi2017pointnet} leads to more accurate prediction compared to using a 2D convolutional network.

As shown in Fig.~\ref{fig:network}, given the lidar calibration data, we first convert the range image patch to a mini point cloud (with maximally $hw-1$ points). Instead of directly regressing the pixel range value, which suffers from the uncertainty caused by the multi-modal distribution of attributes (esp. on the object boundaries), we formulate the prediction as an anchor-based classification and anchor-residual regression problem, where valid pixels in the range image patch are the anchors. The deep network predicts which pixel is the closest in value to the bottom right pixel and regresses a residual (it is an overloaded word here; it is different from the residual map in delta encoding) with respect to each anchor pixel.

\paragraph{Temporal Model}
The temporal model extends the intra-frame prediction model by leveraging contexts from both the current scan and the past scan. The point cloud representation (compared to the 2D pixel representation) enables us to unify the input from the past and current scans as we can represent all laser shots in the 4D (spherical plus time) coordinates.

Given the current scan (quantized) range image $I'_{T}$ and the past scan range image $I'_{T-1}$, assume we want to predict the range value of pixel $(i, j)$ in the current scan ($k$-th pixel in the raster scanning order). A naive baseline approach to use temporal data is to take the same neighborhood at that in $I'_{T}$ (in terms of pixel rows and columns) from the last scan $I'_{T-1}$ and concatenate it with the current frame image patch. However, this approach does not take the ego-motion of the lidar sensor into account. As the lidar moves over time, the range image patch with the same rows and columns can correspond to vastly different physical space.

To take sensor poses into consideration, instead of querying pixels of the last frame using the row and column indices, we should query neighbors using 3D points in the global coordinate (Fig.~\ref{fig:network}). However, as we do not know the ground truth range value for the pixel $(i, j)$, we have to approximate the query by using a predicted range (e.g. using the left pixel range or the predicted value from the intra-frame model). Given pixel $(i, j)$'s laser shot angle $(\theta, \alpha)$ and its estimated range $\hat{r}$, we get a point in the global coordinate, following Sec.~\ref{sec:problem}. Then given the points from last frame in the global coordinate, we can directly query neighbors in the 3D space (using KDtrees to accelerate the query). Those neighboring points from last frame can then be projected to laser shot $(i, j)$'s spherical coordinate (to the points in the sensor coordinate at the time of the laser shot and then transform to the spherical coordinate), to obtain extra points as temporal contexts.~\footnote{Strictly, even the pixels/points from the current frame need be re-projected to the sensor coordinate at the time of the shot $(i, j)$. We have this reprojection in our intra-frame model but the impact is small as the sensor moves little between a few pixels.}
This is equivalent to assuming the points from the last frame are static, and we re-scan the scene at the sensor location at the time of the laser shot $(i, j)$. To distinguish the points from the last and current frames, we augment the points with an extra time channel (with 1 indicating the last frame and 0 indicating the current frame).

Note that the reprojected points from the last frame do not directly correspond to the rows and columns of the current frame range image. Considering such input as a point cloud is convenient as we do not require any interpolation to turn the points to the image grid or any predefined neighborhood size for image cropping.


\paragraph{Inference.}
At inference time (for compression), we start from the top left patch of the range image to predict pixel $I'_1$ or $I'_{1, 1}$ and store the residual. This process continues in a raster scanning order to predict pixels $I_{1, 2}, ..., I_{1, W}, I_{2, 1}, ...,I_{i, j}, ...,I_{H, W}$. The residual map (deltas between the prediction and quantized values) of size $H\times W$ would be compressed by the entropy encoder. At decompression time, we run the prediction model in the same raster-scanning order, which takes input as already reconstructed pixels $\{I'_1, .., I'_{k-1}\}$, predicts the next pixel value $\hat{I}_k$ and then reconstruct the pixel from saved residual as $I'_k = \hat{I}_k + \delta_k$, where $\delta_k$ is the stored delta of pixel $k = (i-1)W + j$. This process can be parallelized by dividing the input range image into blocks and run the inference in parallel for each block (discussed in the supplementary).

\subsection{Entropy Encoding}
After the predictive delta encoding, we get a residual map/array of the range image. An entropy encoder is used to leverage the sparsity pattern in the residual map to compress it. Given an accurate prediction model, most of the residuals would be zero.
We adopt two methods to entropy encode the residuals. In practice, we select the entropy encoder with the highest compression rates depending on the quantization rates and the predictor.

The first method is to represent the residuals using a sparse representation, with the values of the nonzero residuals and their indices in the array, which can then be arithmetically encoded to further reduce its size.
The second method is to represent the residuals using run-length encoding, which achieves better compression rates when the residuals are not very sparse, i.e., when quantization step is small. After obtaining the run-length representation, we use LZMA compressor to further reduce its size.

\label{section:approach:entropy}


%% file: tex/exp.tex

In this section, we first introduce the datasets and the metrics in Sec.~\ref{sec:exp:dataset}. Then we report compression results compared with strong baselines and prior art methods in Sec.~\ref{sec:exp:results} both quantitatively and qualitatively. We further evaluate the impact of compressed data to downstream perception tasks (3D detection of vehicles and pedestrians) in Sec.~\ref{sec:exp:detection_results}. Finally, we provide extensive analysis experiments to validate our design choices in Sec.~\ref{sec:exp:analysis}.

\subsection{Dataset and Metrics}
\label{sec:exp:dataset}

\paragraph{Waymo Open Dataset (WOD)~\cite{sun2020scalability}} WOD is the main dataset we experiment with, as it provides rich lidar calibration data and full sensor poses. WOD includes a total number of 1,150 sequences with 798 for training and 202 for validation. Each sequence lasts around 20 seconds with a sampling frequency of 10Hz.
A 64-beam lidar is used, providing range images of 64 rows and 2,650 columns, with provided lidar calibration metadata (beam inclination angles). The range channel is cropped to 75m, and each raw range value is stored as a 32-bit float in default. We use the training set to train our deep model and evaluate on the validation set. Only the first return range images are used in our experiments.

\paragraph{SemanticKITTI~\cite{behley2019semantickitti}} We also evaluate our method on SemanticKITTI (which enhances KITTI~\cite{geiger2013vision} with semantic labels) to compare with prior art methods OctSqueeze~\cite{huang2020octsqueeze} and MuSCLE~\cite{biswas2020muscle} (since they do not release code, we cannot compare with them on the WOD). We directly apply the WOD trained model on SemanticKITTI test split (sequence 11-21). However, as KITTI only released the point cloud data but not the the raw range images nor the sensor poses, we have to refer to the manual of the Velodyne lidar~\cite{HDL-64E} used by KITTI to convert a point cloud to the spherical coordinate to get a \emph{pseudo} range image with 64 rows and 2,088 columns. For our method, we compress the pseudo range images and do not additionally store the azimuth and elevation of the pixels, as their storage in actual Velodyne range images are negligible (elevations are known and azimuths can be compressed to less than 1Kb per frame~\cite{tu2016compressing}).

\paragraph{Metrics} Following previous works~\cite{graziosi2020overview, huang2020octsqueeze,biswas2020muscle}, we use two geometric metrics to evaluate the reconstruction quality of the compressed point cloud data: point-to-point Chamfer distance and point-to-plane peak signal-to-noise ratio (PSNR). We report these metrics as a function of bitrates i.e., the average number of bits to store one lidar point.

The point-to-point Chamfer distance $\text{CD}_{sym}$ measures the average point distances between two point clouds (smaller the better). For a given point cloud $P=\{p_i\}_{i=1,...N}$ and the reconstructed point cloud $\hat{P}=\{\hat{p}_j\}_{j=1,...M}$:

\begin{equation}
    \text{CD}(P, \hat{P}) = \frac{1}{\vert P \vert} \sum_i \min_j \lVert p_i - \hat{p}_j \lVert_{2}
\end{equation}
\begin{equation}
    \text{CD}_{sym}(P, \hat{P}) = \text{max}\{\text{CD}(P, \hat{P}), \text{CD}(\hat{P}, P)\}
\end{equation}

The second metric, the peak signal-to-noise ratio (PSNR)~\cite{tian2017geometric} (the larger the better), measures the ratio between the ``resolution'' of the point cloud $r$ and the average point-to-plane error between the original point cloud $P$ and the reconstructed point cloud $\hat{P}$:

\begin{equation}
    \text{PSNR}(P, \hat{P})=10\log_{10} \frac{r^2}{\max\{\text{MSE}(P, \hat{P}),\text{MSE}(\hat{P}, P)\}}
\end{equation}
where $\text{MSE}(P, \hat{P}) = \frac{1}{\vert P \vert} \sum_i((p_i - \hat{p}_i) \cdot n_i)^2$ is the point-to-plane distance, $\hat{p}_i$ is the closest point in $\hat{P}$ to $p_i$,  $r=\max_{p_i\in P}\min_{j\not=i}\lVert p_i - p_j\lVert_{2}$ is the intrinsic resolution of the original point cloud. We estimate the normal $n_i$ using Open3D~\cite{zhou2018open3d} with $k = 12$ for k nearest neighbor.

\subsection{Compression Results}
\label{sec:exp:results}
In this section, we compare our methods with competitive baselines as well as prior art lidar data compression methods.
We focus on compressing the range channel or the 3D coordinates of the points as it is the most studied attribute among the others (intensity, elongation) and some of the methods in comparison do not support compressing other attributes. See supplementary material for more results on compressing the other channels. We adjust the quantization precision of the range images to achieve different compression rates (bits per point) of our method. 

\paragraph{Baselines:} \textbf{G-PCC}~\cite{graziosi2020overview} is a point cloud compression method proposed by the MPEG, using octrees. \textbf{Draco}~\cite{Draco} is a popular point cloud compression algorithm based on Kdtrees proposed by Google. We also compare with two prior art deep model based methods~\footnote{There is another deep net based work VoxelContextNet~\cite{que2021voxelcontext}, yet as they did not release code nor the detailed definition of the evaluation metrics, we could not compare with them.}: \textbf{OctSqueeze}~\cite{huang2020octsqueeze} is a octree-based method that uses a neural network to predict the next-level symbol of the octree; \textbf{MuSCLE}~\cite{biswas2020muscle} further strengthens OctSqueeze by leveraging multi-sweep (temporal) data for the octree prediction. In terms of range image representation, we compare with \textbf{PNG} (intra-frame) as well as \textbf{HEVC} (a video compression standard) on top of PNG for temporal range image compression. For the PNG compression, the range is coded with 16 bits with a varying scaling factor to control the distortion/compression rate. We also compare with \textbf{Cluster}~\cite{sun2019novel}, a range image-based lidar data compression algorithm with a pipeline of segmentation, clustering, 3D-HEVC encoding and ground prediction. Besides, supplementary provides a further experiment comparing with an auto-encoder based method on range images (not included here due to its poor performance).

\paragraph{Implementation Details} Our intra-frame prediction model, \textbf{RIDDLE}, takes in a context image patch of size $10 \times 10$  (the bottom right pixel is masked out) and uses a PointNet~\cite{qi2017pointnet} like architecture for the prediction (without the T-Net structure, adapted the output to predict anchor classification and regression). The input to the network is a 3D point cloud in a spherical coordinate with azimuth, elevation relative to the bottom right pixel and the range relative to the mean range of valid context points. Our temporal model, \textbf{RIDDLE-T}, uses the same network architecture as the intra-frame one but takes in an extra 100 points from the last scan (projected to the spherical coordinate of the next pixel). Please see supplementary for more details.

\begin{figure*}[t!]
\centering
\includegraphics[width=0.8\linewidth]{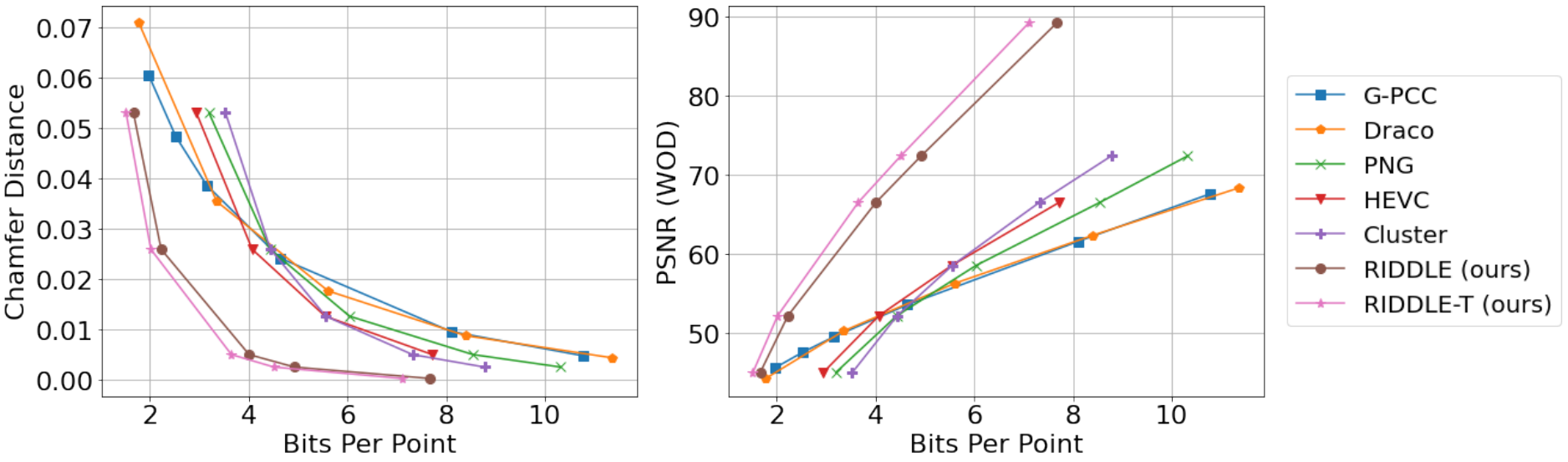}
\caption{\textbf{Evaluation of the compression methods with geometric metrics on the \emph{Waymo Open Dataset val set}.} \emph{Left}: Chamfer distance v.s. bit per point (bbp); \emph{Right}: PSNR v.s. bpp. At a certain bitrate, lower the Chamfer distance or higher the PSNR, better the reconstruction quality.}
\label{fig:wod_geometry}
\end{figure*}

\paragraph{Waymo Open Dataset Results}
We report the bitrate versus reconstruction quality metrics (PSNR, Chamfer distance) of competing methods on all frames from the sequences in the validation set of the Waymo Open Dataset.
As shown in Fig.~\ref{fig:wod_geometry}, our method significantly outperforms prior methods.
At the same Chamfer distance around 0.005, our method reduces the bitrate by more than 65\% compared to G-PCC (from 10.78 bpp to 3.65 bpp). At the bitrate of around 4, our method reduces the distortion (measured by Chamfer distance) by more than 85\%.
Our method also has a larger bitrate improvement over previous methods when the reconstruction quality is higher. This indicates our method has more advantage over baselines when the data quality requirement is higher.

\begin{figure}[t!]
\centering
\includegraphics[width=\linewidth]{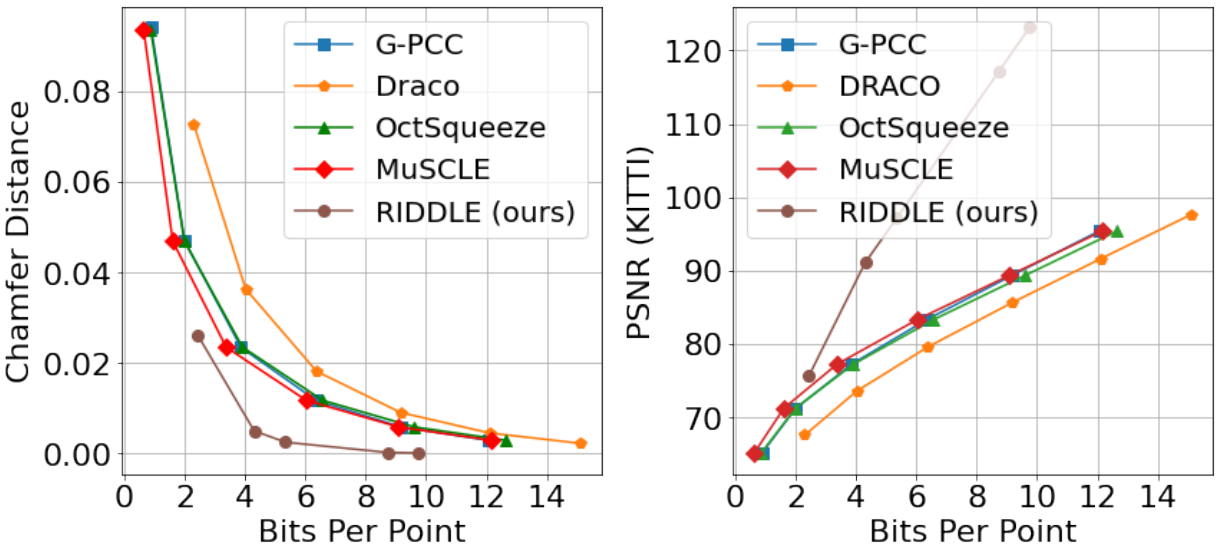}
\caption{\textbf{Evaluation of the compression methods with geometric metrics on the \emph{SemanticKITTI test set}.} We only present our intra-frame model here as the per pixel sensor pose is unavailable in SemanticKITTI.}
\label{fig:kitti_geometry}
\end{figure}

\begin{figure}[t!]
\centering
\includegraphics[width=\linewidth]{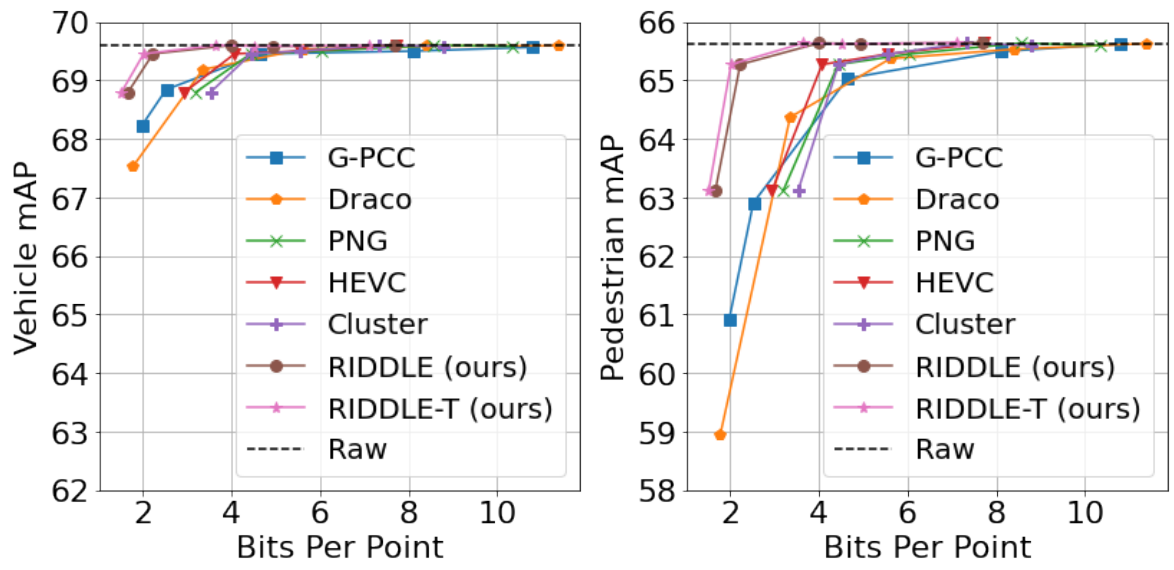}
\caption{\textbf{Impact of lidar data compression to 3D object detection quality on the \emph{Waymo Open Dataset val set}.} We train PointPillars~\cite{lang2019pointpillars} detectors using the raw point clouds (with no compression) from the WOD train set and evaluate them with the compressed point clouds (or point clouds from the compressed range images) on the WOD validation set.}
\label{fig:detection_results}
\end{figure}

\paragraph{SemanticKITTI Results}
Since prior art methods~\cite{huang2020octsqueeze, biswas2020muscle} have not released the code or the compression model, we turn to the SemanticKITTI dataset to compare with them (we got the raw values of the curves reported in the MuSCLE~\cite{biswas2020muscle} paper from the authors).
We apply our model trained on the Waymo Open Dataset directly to the SemanticKITTI lidar point clouds (by creating pseudo range images).

As shown in Fig.~\ref{fig:kitti_geometry}, our method is more than 50\% lower in bitrate (at around 4.3 bpp) with the same Chamfer distance at around 0.005 compared to all prior art methods, showing significant advantages. This strong lead attributes to our choice of directly compressing the range images as well as the effective deep model.

\begin{figure*}[t!]
\centering
\includegraphics[width=0.88\linewidth]{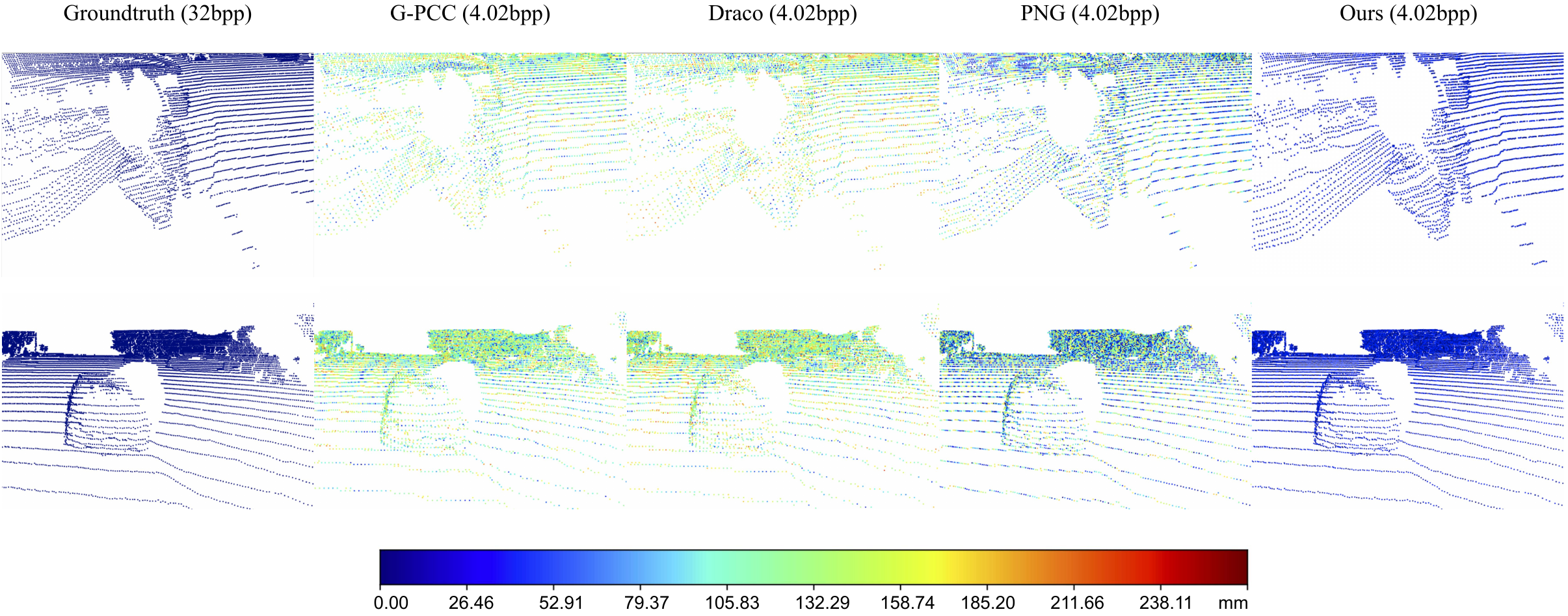}
\caption{\textbf{Visualization of reconstructed point clouds, colored by per point Chamfer distance} (error bar colormap on the bottom). From left to right: raw, G-PCC, Draco, PNG and RIDDLE (ours). It is clear that our method, under the same bit per point, has mush less distortion. Best viewed in color with zoom in.}
\label{fig:qualitative}
\end{figure*}

\paragraph{Qualitative results.}
In Fig.~\ref{fig:qualitative}, we show the reconstructed lidar point clouds from our method, Draco and G-PCC. We can see that the point cloud reconstructed from our method remarkably resembles the original point cloud in geometry even when the bitrate is ambitiously set very low, thanks to compressing directly on the range images to keep the point distribution pattern.

\subsection{Impact to Downstream Perception Tasks}
\label{sec:exp:detection_results}
For applications like autonomous driving, we want to understand the impact of lidar data compression to downstream perception tasks such as 3D object detection.
To understand such impact, we trained a widely used PointPillars detector~\cite{lang2019pointpillars} on uncompressed point clouds using the Waymo Open Dataset train set, for the vehicle class and pedestrian class respectively. Detection quality is measured by mean average precision (mAP).

As shown in Fig.~\ref{fig:detection_results}, our method outperforms other competing baselines in maintaining the best mAP with the same bitrate. At the bitrate around 2, our method leads the second best method (G-PCC) by more than 1 point on vehicle detection and 3 points on pedestrian detection. We can also see that pedestrian detection is more sensitive to data distortion probably due to the smaller average object sizes compared to vehicles.

\begin{table*}[t!]
\small
\parbox{.32\linewidth}{
\begin{center}
\begin{tabular}{lc}
\toprule
\multicolumn{1}{l}{model}  &\multicolumn{1}{c}{acc. @0.1m} 
\\ \hline
previous valid value & 54.35 \\
linear interpolation & 54.64 \\
12-layer CNN          &64.62 \\
PointNet (adpated)  &\textbf{65.75} \\
\bottomrule
\end{tabular}
\end{center}
\caption{Effects of prediction models.}
\label{tab:ablation:model}
}
\hfill
\parbox{.32\linewidth}{
\begin{center}
\begin{tabular}{lc}
\toprule
\multicolumn{1}{l}{loss function}  &\multicolumn{1}{c}{  acc. @0.1m} 
\\ \hline
MSE & 59.83\\
MAE & 61.64\\
multi-bin loss & 59.66\\
anchor cls. + reg. &\textbf{65.75} \\
\bottomrule
\end{tabular}
\end{center}
\caption{Effects of loss functions.}
\label{tab:ablation:loss}
}
\hfill
\parbox{.32\linewidth}{
\begin{center}
\begin{tabular}{lc}
\toprule
\multicolumn{1}{l}{temporal context} &\multicolumn{1}{c}{acc. @0.1m} 
\\ \hline
none (intra-frame) & 65.75\\
$10 \times 10$ image & 67.34\\
$100$ knn points & \textbf{69.23}\\
\bottomrule
\\
\end{tabular}
\end{center}
\caption{Effects of temporal input.}
\label{tab:ablation:temp}
}
\end{table*}

\subsection{Analysis Experiments}
\label{sec:exp:analysis}
In this section we ablate our deep model in terms of architecture choice, loss design and temporal context. In order to compare prediction quality independent from the entropy encoder, we use a prediction accuracy as the metrics for ablation studies. The prediciton accuracy (acc.) is defined as the percentage of zero deltas (i.e. perfect prediction under quantization) in the range image residual map, under a specific quantization precision (e.g. $\delta = 0.1$m~\footnote{Note $0.1$m is not that coarse as average point displacement after the quantization is only $2.5$cm}). A prediction $q$ for the quantized range value $p'$ is counted as correct if $|q - p'| < \delta / 2$.
Supplementary provides more analysis related to entropy encoders and model latency.

\paragraph{Effects of predictor choices.}

Table~\ref{tab:ablation:model} compares several architecture choices. The simplest choice is to use the left valid pixel as the prediction to the current pixel: $\hat{I}_{i,j} = I'_{i, j-1}$. Another extension is to use linear interpolation of close-by pixels: $\hat{I}_{i,j} = I'_{i,j-1} + I'_{i-1,j} - I'_{i-1,j-1}$. Note that for both cases, first valid pixel is used in case the nearby one is an empty pixel. We see that deep models can significantly outperform linear models while the point-cloud-based architecture shows a stronger empirical result compared to ConvNet on the image representation.

\paragraph{Effects of loss functions.}
Table~\ref{tab:ablation:loss} compares several loss choices for our model supervision. With direct attribute prediction as a regression problem, we can see using the mean absolute error (MAE, L1 loss) is superior to using the mean squared error (MSE, L2 loss) as it is affected less by the large errors on the object boundaries. Turning the depth regression problem to a multi-bin classification and regression problem (with classification and intra-bin regression for each depth bin of size 1m) does not help much either as shown in the third row. Our proposed formulation (anchor classification with regression) leads to 4.11 points increase in prediction accuracy compared to the second best option of using mean absolute error.

\paragraph{Effects of temporal contexts.}
Table~\ref{tab:ablation:temp} shows the benefits of adding temporal contexts to the prediction model. We see that even the naive concatenation of the image patch of the last frame with the same rows and columns (second row) can already help. A more careful handling of the temporal points by considering sensor poses (as described in Sec.~\ref{sec:approach:network}) leads to more gains of using the temporal data.

%% file: tex/conclusion.tex
With improving lidar sensor resolution and growing data volume, how to efficiently store and transmit lidar data becomes a challenging problem in many 3D applications, such as autonomous driving and augmented reality. To address this challenge, we propose a novel lidar data compression algorithm named RIDDLE (Range Image Deep DeLta Encoding), which combines the succinctness of traditional delta encoding and the expressiveness of deep neural networks, with support of using temporal contexts. Experiments over the Waymo Open Dataset and KITTI show that compared to previous methods, the proposed approach yields significant improvement in the point cloud reconstruction quality and the downstream perception model performance, under the same compression rates.

%% file: tex/appendix.tex

\section{Overview}
In this supplementary, we provide more details of our method, extra analysis experiment results and visualizations. In Sec.~\ref{sec:supp:model}, we describe more details of the deep predictive model, including its network architecture, losses and its training process as well as more explanation of the input data transformations. In Sec.~\ref{sec:supp:analysis}, we provide more analysis experiment results on model latency, effects of the entropy encoder choices and effects of the context sizes. In Sec.~\ref{sec:supp:attributes}, we apply our method to compress lidar data attributes beyond the range values. Finally, in Sec.~\ref{sec:supp:vis}, we provide more visualizations of our model's predictions.

\section{Details of the Predictive Models}
\label{sec:supp:model}
\begin{figure*}[t!]
\centering
\includegraphics[width=0.8\linewidth]{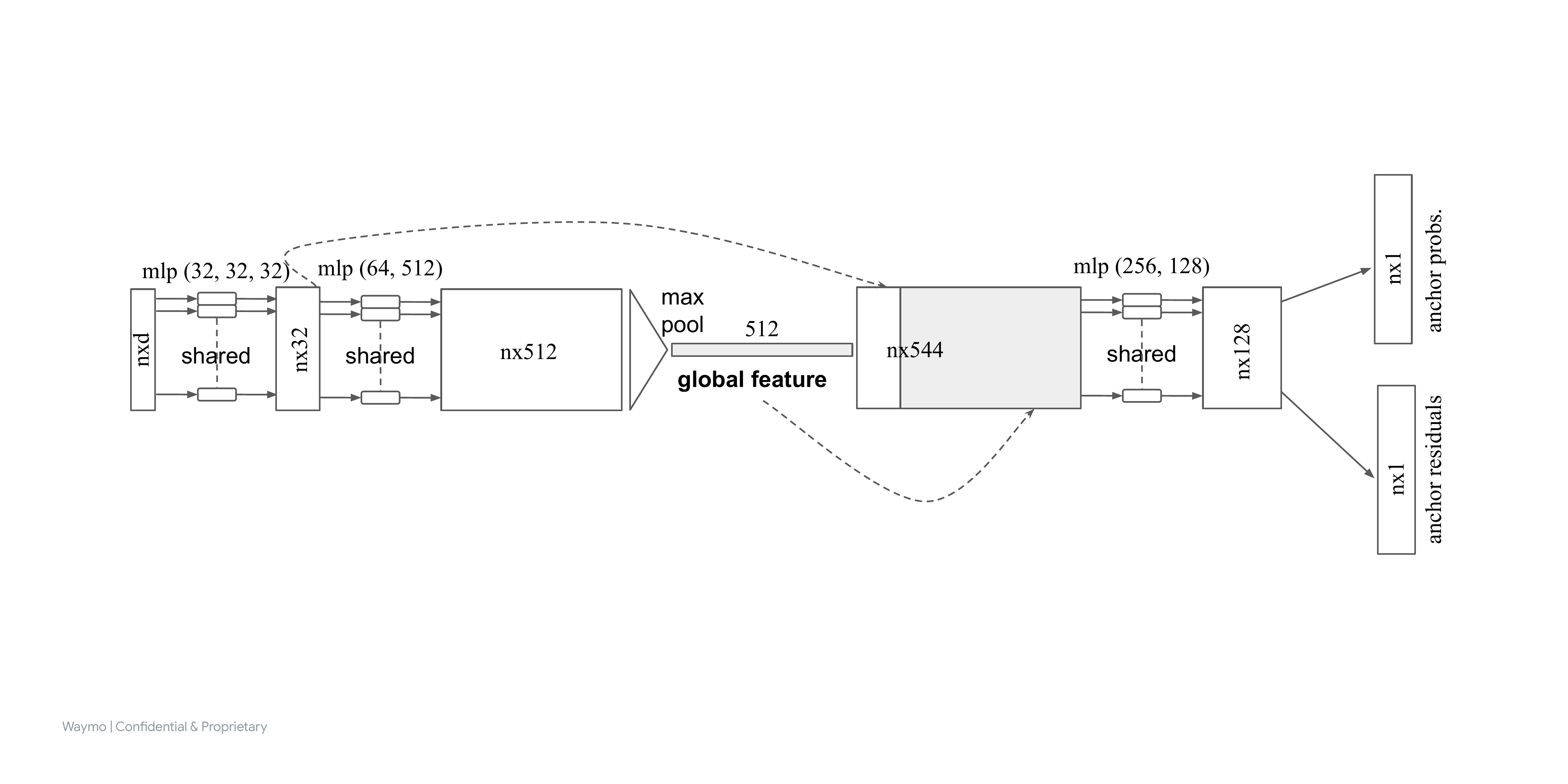}
\caption{\textbf{Deep predictive model architecture.} $n$ is the number of context points, $d$ is the input point dimension ($d=3$ for the intra-prediction model, $d=4$ for the temporal model).}
\label{fig:architecture}
\end{figure*}

\paragraph{Model architecture} For deep prediction model, we adapt the structure of PointNet \cite{qi2017pointnet}. Details of layers are visualized in Fig.~\ref{fig:architecture}. To reduce the latency, the network channel sizes are halved compared to the original architecture and the T-Nets are removed. After concatenation of global and local features, we split the network into anchor-classification branch and residuals branch. Each branch is a MLP with layer sizes [128, 64, \# of anchors], where \# of anchors is 99 for intra-prediction model and 199 for temporal model. For temporal model, We build the KDtrees using neighbors.NearestNeighbors method by Scikit-learn library. We use the left valid depth and up valid depth as estimates to each query 50 points from the last frame as the temporal context.

\paragraph{Loss functions}
At training time, we train deep prediction model end-to-end with the anchor classification and the anchor residual regression loss. We weight the classification loss by a weight.
\begin{equation}
\mathcal{L} = \gamma \mathcal{L}_\text{classification} + \mathcal{L}_\text{regression}
\end{equation}
The classification loss is a cross-entropy loss across $hw - 1$ classes for intra-frame prediction model and $hw - 1 + m$ classes for temporal model. The ground truth class is selected as the index of the pixel with the closest distance to the to-be-predicted pixel. As the input are quantized values, there could be ties. To avoid ties we add a bias term to the distances to favor the pixels that are closer in angles (absolute delta azimuth + absolute delta elevation) to the to-be-predicted pixel. The regression loss is a L1 loss between the predicted residual corresponding to the pixel of the ground truth anchor and the ground truth residual of that pixel.

\paragraph{Training}
We learn the weights of the prediction model by training on the range images from the Waymo Open Dataset train set. We randomly crop the patches of shape $10\times 10$ from the range images and train the deep network with batch size 128 and an Adam optimizer. We use loss weight $\gamma=0.01$. The initial learning rate is 0.00005, and we decay learning rate by 10x at step 1500k and step 3000k. We normalize the range values to $[0, 1]$ by dividing them by 75m. To mimic the same setting in decoding, the training inputs are quantized. The ground truth attribute values are in full precision for more accurate supervision. For pixel locations at
the boundary of the range images, we enforce the same patch size via zero padding.

There are two strategies for inputs quantization. The first strategy is to train different models for different quantization precisions, and for each model we use a fixed quantization precision for the input. The second strategy is using mixed precisions to quantize the input during training. Specifically, we uniformly sample a quantization precision for a given input from 0.0001 to 0.5000 with sample bin size of 0.0001. By the second strategy, we only need to train one model for different compression rates. From our experiments, we observe that the second strategy won't harm the compression rates at individual quantization precision.

\paragraph{Baselines}
For the previous valid value method, we predict the attribute  $I_{i,j}$ at row i, column j to be $\hat{I}_{i,j} = I'_{i,j-1}$, if $I'_{i,j-1}$ is a valid pixel (not void due to empty laser return) else repeatedly decrement j by 1 until the pixel is valid. For the linear interpolation baseline method, we predict $\hat{I}_{i,j} = I'_{i,j-1} + I'_{i-1,j} - I'_{i-1,j-1}$. For the 12-layer CNN method, we adapt a similar structure as ResNet~\cite{he2016deep}. The network is composed of two convolutional layers (channel sizes 64, 32) and 5 residual blocks (channel sizes 32, 32 for each block), and all convolutional layers have filter size 3x3.

\section{More Analysis}
\label{sec:supp:analysis}
\paragraph{Analysis of the model latency.} Table~~\ref{tab:latency} shows the latency comparison between our method and OctSqueeze~\cite{huang2020octsqueeze} and G-PCC from MPEG. To achieve faster decompression speed, during compression, we split a range image into smaller blocks and run compression in parallel on these blocks. During decompression, we decode in parallel on these smaller blocks. Our experiments show that if we split a 64 by 2650 range image into 212 blocks with size 16 by 50, the bitrate would only increase by $0.5\%$, which is nearly negligible. If we split it into 424 blocks with size 16 by 25, the bitrate would increase by $5.13\%$. Table~\ref{tab:latency} shows the latency of our method by splitting into blocks with size 16 by 26 during decoding. We benchmark our method on NVIDIA Tesla V100 GPU. Our deep model is accelerated by TensorRT with float16 quantization. Operations (entropy encoding) other than model inference is written in C++. The latency of G-PCC is benchmarked on CPU using MPEG's implementation (github.com/MPEGGroup/mpeg-pcc-tmc13). The latency of OctSqueeze (depth 16) is from the original paper \cite{huang2020octsqueeze}. Moreover, we believe further speedup of our method could be achieved by methods like predicting multiple pixels at a time or shared point embedding for streamed prediction.

\begin{table}[t]
\begin{center}
\small
\begin{tabular}{lcc}
\toprule
entropy encoder & quantization & bpp
\\ \hline 
sparse repre. + arithmetic encoding & 0.1m & \textbf{2.28} \\
varints+LZMA & 0.1m & 2.33 \\
huffman encoding & 0.1m &  2.49\\
arithmetic encoding & 0.1m & 2.41\\ \hline
sparse repre. + arithmetic encoding & 0.02m & 4.17 \\
varints+LZMA & 0.02m & \textbf{4.08} \\
huffman encoding & 0.02m &  4.25\\
arithmetic encoding & 0.02m & 4.21\\
\bottomrule
\end{tabular}
\end{center}
\caption{\textbf{Ablation study of the entropy encoders.}}
\label{tab:ablation:entropy}
\end{table}

\begin{figure*}[t!]
\centering
\begin{subfigure}{.25\textwidth}
  \centering
  \includegraphics[width=\linewidth]{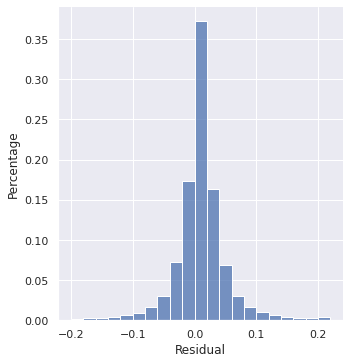}
  \caption{Residuals distribution (0.02m)}
  \label{fig:residualdist002}
\end{subfigure}%
\begin{subfigure}{.25\textwidth}
  \centering
  \includegraphics[width=\linewidth]{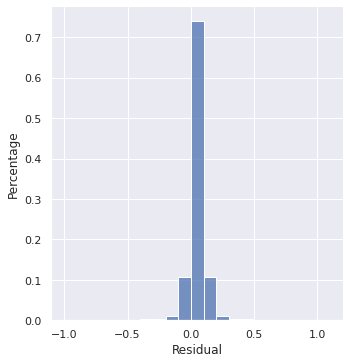}
  \caption{Residuals distribution (0.1m)}
  \label{fig:residualdist01}
\end{subfigure}
\begin{subfigure}{.25\textwidth}
  \centering
  \includegraphics[width=\linewidth]{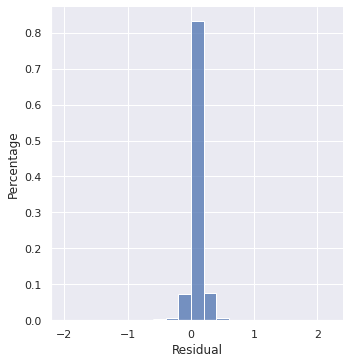}
  \caption{Residuals distribution (0.2m)}
  \label{fig:residualdist02}
\end{subfigure}
\caption{\textbf{Distribution of residuals at different quantization precisions}. Proposed deep prediction model is able to accurately model the joint distributions of range image pixel attributes, resulting in a concentrated distribution of residuals with low entropy.}
\label{fig:appendix:residualdist}
\end{figure*}

\paragraph{Choices of entropy encoder}
After the predictive delta encoding, we get a residual map of the range image. An entropy encoder is used to leverage the sparsity pattern in the residual map to compress it. Given an accurate prediction model, most of the residuals would be zero. In addition, as shown in Fig.~\ref{fig:appendix:residualdist}, larger quantization steps would round more residuals to zero, thus the residuals would become more sparse. 
We adapted two methods to entropy encode the residuals. In practice, we can select the entropy encoder with the highest compression rates depending on the quantization rates and the predictor.

The first method is to represent the residuals using sparse representation. Given an array of residuals, we represent the array with the values of nonzero residuals and their indices in the array. For a long run of sparse residuals, the sparse representation would be quite memory efficient. After obtaining the sparse representation of residuals, we use arithmetic encoding to further reduce its size.

The second method is to represent the residuals using run-length encoding. We first flatten the residual map to a vector and then represent it with the values and the run-length of values. This representation achieves better compression rates when the residuals are not that sparse, i.e. when quantization step size is small. After obtaining the run-length representation, we use LZMA compressor to further reduce its size.

Table.~\ref{tab:ablation:entropy} shows that different entropy encoders have different compression rates of the residuals.
For quantization precision of 0.1m, the residuals are more sparse, and the compression rate of using sparse representation (representing non-zero residuals by specifying their row, col index and the residual values) with arithmetic encoding is higher than varints with LZMA. However, for quantization precision of 0.02m, the compression rate of varints with LZMA is higher, due to the decrease of zero residuals. 

\paragraph{Analysis on input choices.}
Table~\ref{tab:ablation:input} shows how the input choices affect the prediction accuracy. Instead of inputing intra-frame context of 10 by 10 (minus the bottom right one), we can just input the up 9 pixels (row 1), the right 9 pixels (row 2) or smaller context size (row 4). We can see that enlarging the receptive field of input with context from both upper left and upper right can improve the prediction accuracy (row 3 v.s. row 1 and 2). Moreover, including azimuth and inclination as additional input attributes can also improve the predictor (row 4 v.s. row 3) compared to just using the relative row/column indices as input.

\paragraph{Generalization of the method.}
When we apply the deep predictive model trained on 64-beam frames on the Waymo Open Dataset directly to the subsampled 32-beam frames, it achieves 2.55 bpp at 0.1m depth precision (only slightly
larger than 2.23 bpp on 64-beams). In addition,
the compressor trained on WOD can apply well in KITTI
(Fig. 5), which shows the generalization of it.

\paragraph{Comparison with LASzip.}
Benchmarked on Waymo Open Dataset, LASzip~\cite{las} has 67.6
PSNR and 0.0048 Chamfer Distance at 10.62 bpp. Our
method achieves 72.39 PSNR and 0.0026 Chamfer Distance
at 4.51 bpp, which clearly outperforms it.

\begin{table}[t!]
\small
\begin{center}
\begin{tabular}{ccc}
\toprule
method  & compressing (ms)
& decompressing (ms) 
\\ \hline
G-PCC~\cite{graziosi2020overview} & 1594.5 & 1052.1\\
OctSqueeze~\cite{huang2020octsqueeze} & 106.0 & 902.3 \\
RIDDLE (ours) & 532.51  & 966.3 \\
\bottomrule
\end{tabular}
\end{center}
\caption{\textbf{Latencies of lidar data compression methods.} Note the G-PCC is evaluated using CPU on the Waymo Open Dataset (WOD). OctSqueeze is evaluated on the KITTI dataset (with a similar range image resolution to WOD) and our method is evaluated on the WOD. Both OctSqueeze and our method use GPU for model inference.}
\label{tab:latency}
\end{table}

\begin{table}[t!]
\small
\begin{center}
\begin{tabular}{ccc}
\toprule
preprocessing (ms) & network (ms)
& entropy encoding (ms) 
\\ \hline
16.23 & 487.4 & 28.88\\
\bottomrule
\end{tabular}
\end{center}
\caption{\textbf{Breakdown of encoding time of RIDDLE.} Preprocessing includes the time to compute and compress a binary mask indicating whether a pixel is a valid return in the range image. Network refers to the model prediction time. Entropy encoding refers to the time used by entropy encoder.}
\label{tab:latency}
\end{table}

\section{Compression of More Attributes}
\label{sec:supp:attributes}
Since a lidar point cloud may contain additional attributes (e.g. intensity, elongation) than the range values, in this section we show how much we can compress the other attributes than ranges. We train a network to take in multi-channel range images and output multi-channel prediction. Specifically, we train a network on the Waymo Open dataset, which contains 3 channels for each point: range, intensity and elongation. The network is modified to have 3 anchor-classification branches and 3 residuals branches for 3 attributes. From Table~\ref{tab:ablation:attr} row 1-4, we can see that a quantization precision 0.02m for range, or 0.1 for intensity, or 0.1 for elongation have similiar effect on the object detector. With those quantization precisions for each attributes, range values account for most of the storage cost (4.04 bpp) compared to the other two (0.88 bpp for intensity and 0.78 bpp for elongation).




\begin{table}[t!]
\small
\begin{center}
\begin{tabular}{ccc}
\toprule
context size & input format & acc. @0.1m
\\ \hline
10 x  1 & ($\Delta$azimuth, $\Delta$inclination, depth)  & 37.53\\
1 ~ x 10 & ($\Delta$azimuth, $\Delta$inclination, depth) & 60.00  \\
5 ~ x 10 & ($\Delta$row index, $\Delta$col index, depth)  & 65.02 \\
5 ~ x 10 & ($\Delta$azimuth, $\Delta$inclination, depth)  & 65.21 \\ 
10 x 10 & ($\Delta$azimuth, $\Delta$inclination, depth)  & \textbf{65.75}\\
\bottomrule
\end{tabular}
\end{center}
\caption{\textbf{Effects of context size and input format.} We used the intra-frame model for this evaluation.}
\label{tab:ablation:input}
\end{table}

\begin{table*}[t!]
\begin{center}
\small
\begin{tabular}{ccc|ccc|c|cc}
\toprule
\multicolumn{3}{c|}{precision} & \multicolumn{3}{c|}{bit per point} & \multirow{2}{*}{total bpp} & \multirow{2}{*}{vehicle mAP} & \multirow{2}{*}{pedestrian mAP} \\
range & intensity & elongation & range & intensity & elongation & & & \\ \hline
- & - & - & 32 & 32 & 32 & 96.00 & 69.59 & 65.62 \\
0.02m & - & - & 4.04 & 32 & 32 & 68.04 &69.60 & 65.63\\
- & 0.1 & - & 32 & 0.88 & 32 & 64.88 &69.59 & 65.84\\
- & - & 0.1 & 32 &  32 & 0.78 & 64.78 & 69.59 & 65.62\\ 
0.02m & 0.1 & 0.1 & 4.04 & 0.88 & 0.78 & 5.70 & 69.59 & 65.62\\
\bottomrule
\end{tabular}
\end{center}
\caption{\textbf{Effects of context size and input format.} The uncompressed attribute is saved as 32-bit float numbers. We that quantizing the intensity or elongation to 0.1 or quantizing the range value to 0.02m has little impact on the detection mAPs. At these selected quantization rates, the range channel accounts for most of the bpp (around 70\%).}
\label{tab:ablation:attr}
\end{table*}


\begin{figure*}[t!]
\centering
\includegraphics[width=0.8\linewidth]{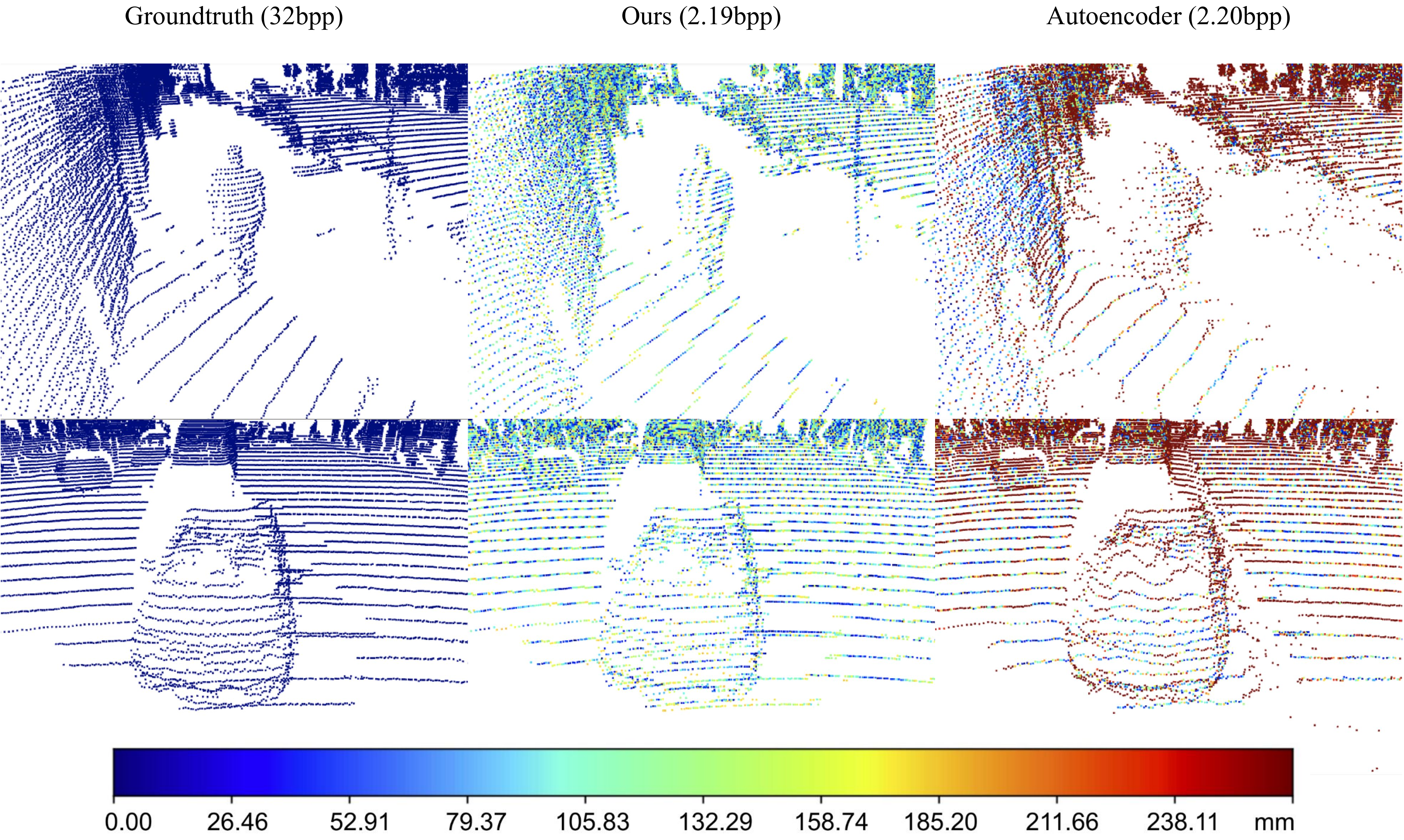}
\caption{\textbf{Visualization of reconstructed point clouds, colored by per point Chamfer distance} (error bar colormap on the bottom). From left to right: raw, RIDDLE (ours) and auto-encoder. It is clear that our method, under the same bit per point, has mush less distortion. Best viewed in color with zoom in.}
\label{fig:autoencoder}
\end{figure*}

\section{More Visualizations}
\label{sec:supp:vis}
\paragraph{The distributions of residuals}
Fig.~\ref{fig:appendix:residualdist} shows the distribution of the range residual maps after our deep delta encoding step. We can see that the larger the quantization interval the more concentrated are the residuals (lower entropy), which explains the lower bitrate after the compression. Note that for a quantization size of 0.1m, more than 70\% of the prediction has zero error compared to the ground truth quantized range image.

\paragraph{Auto-encoder-based compression}
We further compare our method with an auto-encoder-based image compression algorithm \cite{balle2018variational}. $r=\max_{p_i\in P}\min_{j\not=i}\lVert p_i - p_j\lVert_{2}$. The auto-encoder is trained with a learning rate of 0.0001 and an Adam optimizer. The range values are scaled to [0, 1] by 75m instead of by 255 as for RGB images. Fig.~\ref{fig:autoencoder} shows the reconstructed point clouds of the auto-encoder-based method and our method under similar bitrates. The colors of points in the visualizations demonstrate that our method has much better reconstruction quality compared to this auto-encoder baseline. The auto-encoder-based range image compression method does poorly especially at the boundary between foreground points and background points.

%% file: main.bbl
\begin{thebibliography}{10}\itemsep=-1pt

\bibitem{Draco}
Draco.
\newblock \url{https://github.com/google/draco}.
\newblock Accessed: 2021-09-28.

\bibitem{HDL-64E}
Velodyne hdl-64e.
\newblock
  \url{https://gpsolution.oss-cn-beijing.aliyuncs.com/manual/LiDAR/MANUAL\%2CUSERS\%2CHDL-64E_S3.pdf}.
\newblock Accessed: 2021-10-04.

\bibitem{adaprad}
Jae-Kyun Ahn, Kyu-Yul Lee, Jae-Young Sim, and Chang-Su Kim.
\newblock Large-scale 3d point cloud compression using adaptive radial distance
  prediction in hybrid coordinate domains.
\newblock {\em IEEE Journal of Selected Topics in Signal Processing},
  9(3):422--434, 2015.

\bibitem{2017arXiv170201105A}
I. {Armeni}, A. {Sax}, A.~R. {Zamir}, and S. {Savarese}.
\newblock {Joint 2D-3D-Semantic Data for Indoor Scene Understanding}.
\newblock {\em ArXiv e-prints}, Feb. 2017.

\bibitem{balle2016end}
Johannes Ball{\'e}, Valero Laparra, and Eero~P Simoncelli.
\newblock End-to-end optimized image compression.
\newblock {\em arXiv preprint arXiv:1611.01704}, 2016.

\bibitem{balle2018variational}
Johannes Ball{\'e}, David Minnen, Saurabh Singh, Sung~Jin Hwang, and Nick
  Johnston.
\newblock Variational image compression with a scale hyperprior.
\newblock {\em arXiv preprint arXiv:1802.01436}, 2018.

\bibitem{beek2019image}
Peter~van Beek.
\newblock Image-based compression of lidar sensor data.
\newblock {\em Electronic Imaging}, 2019(15):43--1, 2019.

\bibitem{behley2019semantickitti}
Jens Behley, Martin Garbade, Andres Milioto, Jan Quenzel, Sven Behnke, Cyrill
  Stachniss, and Jurgen Gall.
\newblock Semantickitti: A dataset for semantic scene understanding of lidar
  sequences.
\newblock In {\em Proceedings of the IEEE International Conference on Computer
  Vision}, pages 9297--9307, 2019.

\bibitem{biswas2020muscle}
Sourav Biswas, Jerry Liu, Kelvin Wong, Shenlong Wang, and Raquel Urtasun.
\newblock Muscle: Multi sweep compression of lidar using deep entropy models.
\newblock {\em arXiv preprint arXiv:2011.07590}, 2020.

\bibitem{botsch2002efficient}
Mario Botsch, Andreas Wiratanaya, and Leif Kobbelt.
\newblock Efficient high quality rendering of point sampled geometry.
\newblock {\em Rendering Techniques}, 2002:13th, 2002.

\bibitem{dai2017scannet}
Angela Dai, Angel~X Chang, Manolis Savva, Maciej Halber, Thomas Funkhouser, and
  Matthias Nie{\ss}ner.
\newblock Scannet: Richly-annotated 3d reconstructions of indoor scenes.
\newblock In {\em CVPR}, 2017.

\bibitem{devillers2000geometric}
Olivier Devillers and P-M Gandoin.
\newblock Geometric compression for interactive transmission.
\newblock In {\em Proceedings Visualization 2000. VIS 2000 (Cat. No.
  00CH37145)}, pages 319--326. IEEE, 2000.

\bibitem{geiger2013vision}
Andreas Geiger, Philip Lenz, Christoph Stiller, and Raquel Urtasun.
\newblock Vision meets robotics: The kitti dataset.
\newblock {\em The International Journal of Robotics Research},
  32(11):1231--1237, 2013.

\bibitem{graziosi2020overview}
D Graziosi, O Nakagami, S Kuma, A Zaghetto, T Suzuki, and A Tabatabai.
\newblock An overview of ongoing point cloud compression standardization
  activities: video-based (v-pcc) and geometry-based (g-pcc).
\newblock {\em APSIPA Transactions on Signal and Information Processing}, 9,
  2020.

\bibitem{he2016deep}
Kaiming He, Xiangyu Zhang, Shaoqing Ren, and Jian Sun.
\newblock Deep residual learning for image recognition.
\newblock In {\em Proceedings of the IEEE conference on computer vision and
  pattern recognition}, pages 770--778, 2016.

\bibitem{houshiar20153d}
Hamidreza Houshiar and Andreas N{\"u}chter.
\newblock 3d point cloud compression using conventional image compression for
  efficient data transmission.
\newblock In {\em 2015 XXV International Conference on Information,
  Communication and Automation Technologies (ICAT)}, pages 1--8. IEEE, 2015.

\bibitem{huang2020octsqueeze}
Lila Huang, Shenlong Wang, Kelvin Wong, Jerry Liu, and Raquel Urtasun.
\newblock Octsqueeze: Octree-structured entropy model for lidar compression.
\newblock In {\em Proceedings of the IEEE/CVF Conference on Computer Vision and
  Pattern Recognition}, pages 1313--1323, 2020.

\bibitem{las}
Martin Isenburg.
\newblock Laszip: lossless compression of lidar data.
\newblock {\em Photogrammetric Engineering and Remote Sensing}, 79, 02 2013.

\bibitem{lang2019pointpillars}
Alex~H Lang, Sourabh Vora, Holger Caesar, Lubing Zhou, Jiong Yang, and Oscar
  Beijbom.
\newblock Pointpillars: Fast encoders for object detection from point clouds.
\newblock In {\em CVPR}, 2019.

\bibitem{ma2019image}
Siwei Ma, Xinfeng Zhang, Chuanmin Jia, Zhenghui Zhao, Shiqi Wang, and Shanshe
  Wang.
\newblock Image and video compression with neural networks: A review.
\newblock {\em IEEE Transactions on Circuits and Systems for Video Technology},
  30(6):1683--1698, 2019.

\bibitem{mentzer2019practical}
Fabian Mentzer, Eirikur Agustsson, Michael Tschannen, Radu Timofte, and Luc~Van
  Gool.
\newblock Practical full resolution learned lossless image compression.
\newblock In {\em Proceedings of the IEEE/CVF conference on computer vision and
  pattern recognition}, pages 10629--10638, 2019.

\bibitem{6943095}
Fabrizio Nenci, Luciano Spinello, and Cyrill Stachniss.
\newblock Effective compression of range data streams for remote robot
  operations using h.264.
\newblock In {\em 2014 IEEE/RSJ International Conference on Intelligent Robots
  and Systems}, pages 3794--3799, 2014.

\bibitem{oord2016conditional}
Aaron van~den Oord, Nal Kalchbrenner, Oriol Vinyals, Lasse Espeholt, Alex
  Graves, and Koray Kavukcuoglu.
\newblock Conditional image generation with pixelcnn decoders.
\newblock {\em arXiv preprint arXiv:1606.05328}, 2016.

\bibitem{qi2017pointnet}
Charles~R Qi, Hao Su, Kaichun Mo, and Leonidas~J Guibas.
\newblock Pointnet: Deep learning on point sets for 3d classification and
  segmentation.
\newblock In {\em Proceedings of the IEEE conference on computer vision and
  pattern recognition}, pages 652--660, 2017.

\bibitem{que2021voxelcontext}
Zizheng Que, Guo Lu, and Dong Xu.
\newblock Voxelcontext-net: An octree based framework for point cloud
  compression.
\newblock In {\em Proceedings of the IEEE/CVF Conference on Computer Vision and
  Pattern Recognition}, pages 6042--6051, 2021.

\bibitem{schnabel2006octree}
Ruwen Schnabel and Reinhard Klein.
\newblock Octree-based point-cloud compression.
\newblock In {\em PBG@ SIGGRAPH}, pages 111--120, 2006.

\bibitem{song2015sun}
Shuran Song, Samuel~P Lichtenberg, and Jianxiong Xiao.
\newblock Sun rgb-d: A rgb-d scene understanding benchmark suite.
\newblock In {\em CVPR}, 2015.

\bibitem{sun2020scalability}
Pei Sun, Henrik Kretzschmar, Xerxes Dotiwalla, Aurelien Chouard, Vijaysai
  Patnaik, Paul Tsui, James Guo, Yin Zhou, Yuning Chai, Benjamin Caine, et~al.
\newblock Scalability in perception for autonomous driving: Waymo open dataset.
\newblock In {\em Proceedings of the IEEE/CVF Conference on Computer Vision and
  Pattern Recognition}, pages 2446--2454, 2020.

\bibitem{sun2019novel}
Xuebin Sun, Han Ma, Yuxiang Sun, and Ming Liu.
\newblock A novel point cloud compression algorithm based on clustering.
\newblock {\em IEEE Robotics and Automation Letters}, 4(2):2132--2139, 2019.

\bibitem{tian2017geometric}
Dong Tian, Hideaki Ochimizu, Chen Feng, Robert Cohen, and Anthony Vetro.
\newblock Geometric distortion metrics for point cloud compression.
\newblock In {\em 2017 IEEE International Conference on Image Processing
  (ICIP)}, pages 3460--3464. IEEE, 2017.

\bibitem{toderici2017full}
George Toderici, Damien Vincent, Nick Johnston, Sung Jin~Hwang, David Minnen,
  Joel Shor, and Michele Covell.
\newblock Full resolution image compression with recurrent neural networks.
\newblock In {\em Proceedings of the IEEE Conference on Computer Vision and
  Pattern Recognition}, pages 5306--5314, 2017.

\bibitem{townsend2019hilloc}
James Townsend, Thomas Bird, Julius Kunze, and David Barber.
\newblock Hilloc: Lossless image compression with hierarchical latent variable
  models.
\newblock {\em arXiv preprint arXiv:1912.09953}, 2019.

\bibitem{tu2019point}
Chenxi Tu, Eijiro Takeuchi, Alexander Carballo, and Kazuya Takeda.
\newblock Point cloud compression for 3d lidar sensor using recurrent neural
  network with residual blocks.
\newblock In {\em 2019 International Conference on Robotics and Automation
  (ICRA)}, pages 3274--3280. IEEE, 2019.

\bibitem{tu2016compressing}
Chenxi Tu, Eijiro Takeuchi, Chiyomi Miyajima, and Kazuya Takeda.
\newblock Compressing continuous point cloud data using image compression
  methods.
\newblock In {\em 2016 IEEE 19th International Conference on Intelligent
  Transportation Systems (ITSC)}, pages 1712--1719. IEEE, 2016.

\bibitem{van2016pixel}
Aaron Van~Oord, Nal Kalchbrenner, and Koray Kavukcuoglu.
\newblock Pixel recurrent neural networks.
\newblock In {\em International Conference on Machine Learning}, pages
  1747--1756. PMLR, 2016.

\bibitem{wiesmann2021deep}
Louis Wiesmann, Andres Milioto, Xieyuanli Chen, Cyrill Stachniss, and Jens
  Behley.
\newblock Deep compression for dense point cloud maps.
\newblock {\em IEEE Robotics and Automation Letters}, 6(2):2060--2067, 2021.

\bibitem{yan2019deep}
Wei Yan, Shan Liu, Thomas~H Li, Zhu Li, Ge Li, et~al.
\newblock Deep autoencoder-based lossy geometry compression for point clouds.
\newblock {\em arXiv preprint arXiv:1905.03691}, 2019.

\bibitem{zhou2018open3d}
Qian-Yi Zhou, Jaesik Park, and Vladlen Koltun.
\newblock Open3d: A modern library for 3d data processing.
\newblock {\em arXiv preprint arXiv:1801.09847}, 2018.

\end{thebibliography}
